\documentclass[12pt,thmsa,a4paper,notitlepage]{article}
\usepackage{sw20lart}

\input{tcilatex}

\begin{document}

\author{Smain BALASKA\thanks{%
e-mail : sbalaska@yahoo.com , balaska.smain@univ-oran.dz} and Toufik SAHABI%
\thanks{%
e-mail : sahabitoufik@yahoo.fr } \and \textit{Laboratoire de physique th\'{e}%
orique d'Oran. } \and \textit{D\'{e}partement de physique. Universit\'{e}
d'Oran Es-S\'{e}nia .} \and \textit{31100 Es-S\'{e}nia. ALGERIA}}
\title{The boundary states and correlation functions of the tricritical
Ising model from the Coulomb-gas formalism$.$}
\maketitle

\begin{abstract}
We consider the minimal conformal model describing the tricritical Ising
model on the disk and on the upper half plane. Using the coulomb-gas
formalism we determine its consistents boundary states as well as its
1-point and 2-point correlation functions.

\end{abstract}

\section{ Introduction}

The basic concepts and techniques of BCFT were introduced first by J. L.
Cardy. He studied in \cite{car2} how to restrict the operator content by
imposing the boundary conditions. He has obtained a classification of the
boundary states and has developed a method to calculate the boundary
correlation functions (\cite{car1} , \cite{car3} and \cite{car4}).

An alternative approach to obtain correlation functions for Cardy's boundary
states is developed in \cite{shin1} and \cite{shin2}. It's a reformulation
of the known Coulomb-gas formalism for theories with boundaries. Instead of
using the contour integration technique (as this was done in \cite{schultze}%
) the author constructs the representation of the boundary states on the
charged bosonic Fock space and uses the charge neutrality conditions to
obtain the correct linear combination of the conformal blocks in the
expression of the correlation functions. This formalism provides a method
for calculating correlation functions in the case of minimal conformal field
theories without having to solve differential equations.

Using this formalism we study the particular case of the minimal conformal
model $(A_{4},A_{3})$ on the upper half plane. The boundary one-point and
two-points correlators of this model are calculated. The supersymmetric
version of the boundary tricritical Ising field theory is studied in \cite%
{d. friedan},\cite{Nepomechie1} and \cite{Nepomechie2}.

The paper is organised as follows. In section 2, we start by considering the
general case of a conformal theory defined on the upper half plane and study
the consequences of the existence of boundaries on the conformal algebra and
on the structure of its Hilbert space. In section 3, using the Cardy's
fusion method \cite{car1} we derive a classification of the conformal
theories one can define on the bounded geometry in terms of the boundary
conditions. In section 4 we determine the consistent boudary states as well
as the partition functions for the particular case of the minimal model $%
(A_{4},A_{3})$.

In section 5, we give the essential of the method used in \cite{shin1} and 
\cite{shin2} and we apply it to determine the one-point as well as the
two-point correlators of the tricritical Ising model on the disk and the
upper half plane.

\section{A CFT on the Upper Half plane}

Let us start with a CFT on the upper half plane (UHP). We define $z=x+iy$
and we consider a theory defined on the region $\func{Im}(z)\geq 0.$

In this case only real analytic changes of coordinates, with

\begin{equation}
\varepsilon (z)=\overline{\varepsilon }(\overline{z})\mathcal{j}_{z=%
\overline{z}\in R}  \label{1}
\end{equation}
are allowed. The energy momentum verify

\begin{equation}
T(z)=\overline{T}(\overline{z})\mathcal{j}_{\text{real axis}}  \label{2}
\end{equation}%
so that there is no momentum flow across the boundary. There is thus only
one copy of the Virasoro algebra 
\begin{equation}
\overline{L}_{-n}=L_{n}\mathcal{j}_{z=\overline{z}}  \label{3}
\end{equation}

Boundary conditions labelled by $\alpha $ and $\beta $ (for $\func{Re}%
(z)\prec 0$ and $\func{Re}(z)\succ 0$) are assigned to fields on the
boundary. The UHP can conformally mapped on an infinite horizontal strip of
width $L$ by $w=\frac{L}{\pi }\ln z$

In both geometries, the system is described by a Hilbert space of states $%
\mathcal{H}_{\alpha \mathcal{j}\beta }$ which decomposes on representations $%
V_{i}$ of the Virasoro algebra according to 
\begin{equation}
\mathcal{H}_{\alpha \mathcal{j}\beta }=\oplus \text{ }n_{i\alpha }^{\quad
\beta }V_{i}  \label{4}
\end{equation}%
where the integers $n_{i\alpha }^{\quad \beta }$ are the multiplicities of
the representations.

On the strip, the Hamiltonian is the translation operator in the direction $%
\func{Re}(w).$ It can be written in the UHP as 

\begin{equation}
H_{\alpha \beta }=\frac{2\pi }{L}(L_{0}-\frac{c}{24})  \label{5}
\end{equation}

To determine the operator content of the theory, we need to determine the
possible boundary conditions $\alpha ,\beta $ on the UHP and also the
associated multiplicities $n_{i\alpha }^{\quad \beta }$

\section{The physical boundary states and the partition functions}

We proceed as in the case of the theories defined on the tore. We take a
semi annular domain in the half plane and identify its edges to make a
cylinder. Equivalently one can consider instead a finite strip. In fact the
semi annular domain comprised between the semi-circles of radius $1$ and $e^{%
\frac{\pi T}{L}}$ (in the $z$ UHP) can be mapped in the $w$ plane into the
segment $0\leq \func{Re}(w)\leq T$ of the strip of width $L$ by the
transformation $w(z)=\frac{L}{\pi }\ln (z)$. In the same way the latter can
also be transformed into the annulus of radius $1$ and $e^{\frac{2\pi L}{T}}=%
\widetilde{q}^{-1/2}$ by the transformation $\zeta (w)=\exp (-2i\pi w/T)$
(with $\widetilde{q}=e^{2\pi i\widetilde{\tau }},\widetilde{\tau }=2iT/L$ ).

The annulus in the $\zeta $ plane is equivalent to a finite cylinder of
length $T$ and circumference $L.$ When considering such geometry one is
allowed to use the familiar energy momentum tensor of the full plane without
modification. Then radial quantization is allowed and the conformal
invariance condition (\ref{2}) on the quantum states $\left\vert \alpha
\right\rangle $ is \cite{car1}

\begin{equation}
(L_{n}^{\zeta }-\overline{L}_{-n}^{\zeta })\left| \alpha \right\rangle =0
\label{6}
\end{equation}

There is a basis of states, which are solution of this linear system of
boundary conditions. It is the basis of Ishibashi states \cite{ishi}. There
exists an independent Ishibashi states (noted $\mathcal{j}j\rangle \rangle $%
) solution of (\ref{6}) for each representation $V_{j}$ of the algebra. The
equation (\ref{6}) being linear, any linear combination of the Ishibashi
states is also a solution. To obtain the combinations corresponding to the
physical boundary states, one uses the Cardy's fusion method \cite{car1}
(for review see also \cite{z1}, \cite{z2})

It consists in calculating the partition function $\mathcal{Z}_{\alpha 
\mathcal{j}\beta }$ in two different ways. First as resulting from the
evolution between the boundary states $\langle \alpha \mathcal{j}$ and $%
\mathcal{j}\beta \rangle $

\begin{eqnarray}
\mathcal{Z}_{\alpha \mathcal{j}\beta } &=&\left\langle \alpha \left|
e^{-TH}\right| \beta \right\rangle  \nonumber \\
&=&\left\langle \alpha \left| \widetilde{q}^{\frac{1}{2}(L_{0}+\overline{L}%
_{0}-\frac{c}{12})}\right| \beta \right\rangle  \label{7}
\end{eqnarray}
where $H=\frac{2\pi }{L}(L_{0}+\overline{L}_{0}-\frac{c}{12})$ and $%
\widetilde{q}=\exp (-4\pi \frac{T}{L})$

On the other hand using the decomposition (\ref{4}) of the hilbert space $%
\mathcal{H}_{\alpha \mathcal{j}\beta }$ one obtain

\begin{equation}
\mathcal{Z}_{\alpha \mathcal{j}\beta }(q)=\sum_{j}n_{j\alpha }^{\text{ ~}%
\beta }\chi _{j}(q)  \label{partion function1}
\end{equation}%
where $\chi _{j}(q)$ is the character of the representation $V_{j}$ and $%
q=e^{-\pi L/T}$

Expanding the physical boundary states as

\begin{equation}
\left| \alpha \right\rangle =\sum_{j}\frac{a_{\alpha j}}{\sqrt{S_{1j}}}%
\mathcal{j}j\rangle \rangle  \label{equation a}
\end{equation}
$S_{ij}$ being the element of the matrix $S$ of the modular group and
equating (\ref{7}) and (\ref{partion function1}), we obtain 
\begin{equation}
\sum_{j}n_{j\alpha }^{\text{ ~}\beta }\chi _{j}(q)=\sum_{jj^{\prime
}}\langle \alpha \mathcal{j}j\rangle \rangle \langle \langle j\mathcal{j}%
\widetilde{q}^{\frac{1}{2}(L_{0}+\overline{L}_{0}-\frac{c}{12})}\mathcal{j}%
j^{\prime }\rangle \rangle \langle \langle j^{\prime }\mathcal{j}\beta
\rangle  \label{10}
\end{equation}

The Ishibashi states are so that 
\begin{equation}
\langle \langle j\mathcal{j}\widetilde{q}^{\frac{1}{2}(L_{0}+\overline{L}%
_{0}-\frac{c}{12})}\mathcal{j}j^{\prime }\rangle \rangle =\delta
_{jj^{\prime }}\chi _{j}(\widetilde{q})  \label{12}
\end{equation}%
then (\ref{10}) becomes

\begin{equation}
\sum_{j}n_{j\alpha }^{\text{ ~}\beta }\chi _{j}(q)=\sum_{j}\frac{a_{\alpha
j}^{*}a_{\beta j}}{S_{1j}}\chi _{j}(\widetilde{q})  \label{13}
\end{equation}

Now performing a modular transformation on the characters 
\begin{equation}
\chi _{j}(\widetilde{q})=\sum_{i}S_{ij}\chi _{i}(q)
\end{equation}%
and identifying the coefficients of $\chi _{i}$, one obtains 
\begin{equation}
n_{j\alpha }^{\text{ \quad ~}\beta }=\sum_{i}\frac{S_{ji}a_{\alpha i}^{\ast
}a_{\beta i}}{S_{1i}}  \label{14}
\end{equation}%
To impose the orthonormality condition $\langle \alpha \mathcal{j}\beta
\rangle =\delta _{\alpha \beta }$, the coefficients $a_{\alpha j}$ have to
verify

\begin{equation}
\sum_{j}a_{\alpha j}^{\ast }a_{\beta j}=\delta _{\alpha \beta }  \label{15}
\end{equation}
As a solution, one can take (recall that the matrix $S$ is unitary)

\begin{equation}
a_{ij}=S_{ij}  \label{16}
\end{equation}
so that finally we have for the physical (or what we also call consistent)
boundary states

\begin{equation}
\left| \alpha \right\rangle =\sum_{j}\frac{S_{\alpha j}}{\sqrt{S_{1j}}}%
\mathcal{j}j\rangle \rangle  \label{17}
\end{equation}
and for the multiplicities

\begin{equation}
n_{j\alpha }^{\text{ \quad ~}\beta }=\sum_{i}\frac{S_{ji}S_{\alpha
j}^{*}S_{\beta j}}{S_{1i}}  \label{18}
\end{equation}

This equation is no more than the Verlinde formula \cite{ver}. It gives the
fusion multiplicities appearing into the decomposition of the fusion of two
representations of the algebra in terms of the elements of the $S$ matrix.

Thus the multiplicities $n_{i\alpha }^{\text{ \quad ~}\beta }$ are
identified with the coefficients of the fusion algebra

\begin{equation}
n_{i\alpha }^{\text{ \quad ~}\beta }=N_{i\alpha }\text{ }^{\beta }
\label{coefficients de fusion}
\end{equation}

We obtain then a classification of the different conformal field theories on
the upper half plane and their operator contents. The later depends on the
boundary conditions.

\section{The case of the tricritical ising model $(A_{4},A_{3}):$}

The central charge, in this case, is $c=7/10$, and the Kac table contents
six primary fields indexed by the pairs $(r,s)$ given in the set $%
\varepsilon .$ See table (\ref{table 1}) for operator contents of the model. 
\[
\varepsilon =\left\{ 
\begin{array}{c}
(1,1)=(3,4),(1,2)=(3,3),(1,3)=(3,2), \\ 
(1,4)=(3,1),(2,2)=(2,3),(2,4)=(2,1)%
\end{array}
\right\} 
\]

$\medskip $

\begin{table}[h]
\centering      
\begin{tabular}{||c||c||c||}
\hline
$(r,s)$ & Dimension & signe \\ \hline
$(1,1)$ or $(3,4)$ & $0$ & $I$ \\ \hline
$(1,2)$ or $(3,3)$ & $1/10$ & $\varepsilon $ \\ \hline
$(1,3)$ or $(3,2)$ & $3/5$ & $\varepsilon ^{\prime }$ \\ \hline
$(1,4)$ or $(3,1)$ & $3/2$ & $\varepsilon ^{\prime \prime }$ \\ \hline
$(2,2)$ or $(2,3)$ & $3/80$ & $\sigma $ \\ \hline
$(2,4)$ or $(2,1)$ & $7/16$ & $\sigma ^{\prime }$ \\ \hline
\end{tabular}%
\caption{The primary fields of the tricritical Ising model.}
\label{table 1}
\end{table}
The modular matrix $S$ is given by

\begin{equation}
S=\frac{1}{\sqrt{5}}\left( 
\begin{array}{cccccc}
s_{2} & s_{1} & s_{1} & s_{2} & \sqrt{2}s_{1} & \sqrt{2}s_{2} \\ 
s_{1} & -s_{2} & -s_{2} & s_{1} & \sqrt{2}s_{2} & -\sqrt{2}s_{1} \\ 
s_{1} & -s_{2} & -s_{2} & s_{1} & -\sqrt{2}s_{2} & \sqrt{2}s_{1} \\ 
s_{2} & s_{1} & s_{1} & s_{2} & -\sqrt{2}s_{1} & -\sqrt{2}s_{2} \\ 
\sqrt{2}s_{1} & \sqrt{2}s_{2} & -\sqrt{2}s_{2} & -\sqrt{2}s_{1} & 0 & 0 \\ 
\sqrt{2}s_{2} & -\sqrt{2}s_{1} & \sqrt{2}s_{1} & -\sqrt{2}s_{2} & 0 & 0%
\end{array}
\right)  \label{/2}
\end{equation}
with $s_{1}=\sin 2\pi /5,$ and $s_{2}=\sin 4\pi /5$\-.

\subsection{The physical boundary states}

As we already mentioned for the diagonal minimal models of type $%
(A_{p-1},A_{p^{\prime }-1})$, the coefficients $a_{\alpha }^{j}$ introduced
in equation (\ref{equation a}) are equal to the elements of the matrix $S$ 
\cite{z2} . Then the Consistent boundary states for this model are written as

\begin{equation}
\left| (r,s)\right\rangle =\sum_{(r^{\prime },s^{\prime })\in \varepsilon }%
\frac{S_{rs,r^{\prime }s^{\prime }}}{\sqrt{S_{11,r^{\prime }s^{\prime }}}}%
\mathcal{j}(r^{\prime },s^{\prime })\rangle \rangle  \label{/3}
\end{equation}

These are the six states given in the following table

\begin{table}[h]
\begin{tabular}{||c||c||}
The field operator & The Consistent boundary state \\ 
$I$ & $\left\vert I\right\rangle =\left\vert (1,1)\right\rangle =\frac{1}{%
\sqrt[4]{5}}\{\sqrt{s_{2}}\mathcal{j}I\rangle \rangle +\sqrt{s_{1}}\mathcal{j%
}\varepsilon \rangle \rangle +\sqrt{s_{1}}\mathcal{j}\varepsilon ^{\prime
}\rangle \rangle +\sqrt{s_{2}}\mathcal{j}\varepsilon ^{\prime \prime
}\rangle \rangle +$ \\ 
& $\sqrt{\sqrt{2}s_{1}}\mathcal{j}\sigma \rangle \rangle +\sqrt{\sqrt{2}s_{2}%
}\mathcal{j}\sigma ^{\prime }\rangle \rangle \}$ \\ 
$\varepsilon $ & $\left\vert \varepsilon \right\rangle =\left\vert
(1,2)\right\rangle =\frac{1}{\sqrt[4]{5}}\{\frac{s_{1}}{\sqrt{s_{2}}}%
\mathcal{j}I\rangle \rangle -\frac{s_{2}}{\sqrt{s_{1}}}\mathcal{j}%
\varepsilon \rangle \rangle -\frac{s_{2}}{\sqrt{s_{1}}}\mathcal{j}%
\varepsilon ^{\prime }\rangle \rangle +\frac{s_{1}}{\sqrt{s_{2}}}\mathcal{j}%
\varepsilon ^{\prime \prime }\rangle \rangle +$ \\ 
& $\sqrt[4]{2}\frac{s_{2}}{\sqrt{s_{1}}}\mathcal{j}\sigma \rangle \rangle -%
\sqrt[4]{2}\frac{s_{1}}{\sqrt{s_{2}}}\mathcal{j}\sigma ^{\prime }\rangle
\rangle \}$ \\ 
$\varepsilon ^{\prime }$ & $\left\vert \varepsilon ^{\prime }\right\rangle
=\left\vert (1,3)\right\rangle =\frac{1}{\sqrt[4]{5}}\{\frac{s_{1}}{\sqrt{%
s_{2}}}\mathcal{j}I\rangle \rangle -\frac{s_{2}}{\sqrt{s_{1}}}\mathcal{j}%
\varepsilon \rangle \rangle -\frac{s_{2}}{\sqrt{s_{1}}}\mathcal{j}%
\varepsilon ^{\prime }\rangle \rangle +\frac{s_{1}}{\sqrt{s_{2}}}\mathcal{j}%
\varepsilon ^{\prime \prime }\rangle \rangle -$ \\ 
& $\sqrt[4]{2}\frac{s_{2}}{\sqrt{s_{1}}}\mathcal{j}\sigma \rangle \rangle +%
\sqrt[4]{2}\frac{s_{1}}{\sqrt{s_{2}}}\mathcal{j}\sigma ^{\prime }\rangle
\rangle $ \\ 
$\varepsilon ^{\prime \prime }$ & $\left\vert \varepsilon ^{\prime \prime
}\right\rangle =\left\vert (1,4)\right\rangle =\frac{1}{\sqrt[4]{5}}\{\sqrt{%
s_{2}}\mathcal{j}I\rangle \rangle +\sqrt{s_{1}}\mathcal{j}\varepsilon
\rangle \rangle +\sqrt{s_{1}}\mathcal{j}\varepsilon ^{\prime }\rangle
\rangle +\sqrt{s_{2}}\mathcal{j}\varepsilon ^{\prime \prime }\rangle \rangle
-$ \\ 
& $\sqrt{\sqrt{2}s_{1}}\mathcal{j}\sigma \rangle \rangle -\sqrt{\sqrt{2}s_{2}%
}\mathcal{j}\sigma ^{\prime }\rangle \rangle \}$ \\ 
$\sigma $ & $\left\vert \sigma \right\rangle =\left\vert (2,2)\right\rangle =%
\frac{1}{\sqrt[4]{5}}\{\sqrt{2}\frac{s_{1}}{\sqrt{s_{2}}}\mathcal{j}I\rangle
\rangle +\sqrt{2}\frac{s_{2}}{\sqrt{s_{1}}}\mathcal{j}\varepsilon \rangle
\rangle -\sqrt{2}\frac{s_{2}}{\sqrt{s_{1}}}\mathcal{j}\varepsilon ^{\prime
}\rangle \rangle -\sqrt{2}\times $ \\ 
& $\frac{s_{1}}{\sqrt{s_{2}}}\mathcal{j}\varepsilon ^{\prime \prime }\rangle
\rangle \}$ \\ 
$\sigma ^{\prime }$ & $\left\vert \sigma ^{\prime }\right\rangle =\left\vert
(2,4)\right\rangle =\frac{1}{\sqrt[4]{5}}\{\sqrt{2s_{2}}\mathcal{j}I\rangle
\rangle -\sqrt{2s_{1}}\mathcal{j}\varepsilon \rangle \rangle +\sqrt{2s_{1}}%
\mathcal{j}\varepsilon ^{\prime }\rangle \rangle -\sqrt{2s_{2}}\mathcal{j}%
\varepsilon ^{\prime \prime }\rangle \rangle \}$%
\end{tabular}%
\caption{The Consistent boundary states of the tricritical Ising model }
\label{table 2}
\end{table}

\subsection{The fusion rules and the partition functions}

The fusion between the different operators of the model are resumed as
follows

\[
\left( 
\begin{array}{llllll}
I & \varepsilon & \varepsilon ^{\prime } & \varepsilon ^{\prime \prime } & 
\sigma & \sigma ^{\prime } \\ 
\varepsilon & \varepsilon ^{2} & \varepsilon \varepsilon ^{\prime } & 
\varepsilon \varepsilon ^{\prime \prime } & \varepsilon \sigma & \varepsilon
\sigma ^{\prime } \\ 
\varepsilon ^{\prime } & \varepsilon ^{\prime }\varepsilon & \varepsilon
^{\prime \text{ }2} & \varepsilon ^{\prime }\varepsilon ^{\prime \prime } & 
\varepsilon ^{\prime }\sigma & \varepsilon ^{\prime }\sigma ^{\prime } \\ 
\varepsilon ^{\prime \prime } & \varepsilon ^{\prime \prime }\varepsilon & 
\varepsilon ^{\prime \prime }\varepsilon ^{\prime } & \varepsilon ^{\prime
\prime \text{ }2} & \varepsilon ^{\prime \prime }\sigma & \varepsilon
^{\prime \prime }\sigma ^{\prime } \\ 
\sigma & \sigma \varepsilon & \sigma \varepsilon ^{\prime } & \sigma
\varepsilon ^{\prime \prime } & \sigma ^{2} & \sigma \sigma ^{\prime } \\ 
\sigma ^{\prime } & \sigma ^{\prime }\varepsilon & \sigma ^{\prime
}\varepsilon ^{\prime } & \sigma ^{\prime }\varepsilon ^{\prime \prime } & 
\sigma ^{\prime }\sigma & \sigma ^{\prime \text{ }2}%
\end{array}
\right) = 
\]

\[
=\text{ {}}\left( 
\begin{array}{llllll}
1 & 0 & 0 & 0 & 0 & 0 \\ 
0 & 1 & 0 & 0 & 0 & 0 \\ 
0 & 0 & 1 & 0 & 0 & 0 \\ 
0 & 0 & 0 & 1 & 0 & 0 \\ 
0 & 0 & 0 & 0 & 1 & 0 \\ 
0 & 0 & 0 & 0 & 0 & 1%
\end{array}
\right) I+\text{ {}}\left( 
\begin{array}{llllll}
0 & 1 & 0 & 0 & 0 & 0 \\ 
1 & 0 & 1 & 0 & 0 & 0 \\ 
0 & 1 & 0 & 1 & 0 & 0 \\ 
0 & 0 & 1 & 0 & 0 & 0 \\ 
0 & 0 & 0 & 0 & 1 & 1 \\ 
0 & 0 & 0 & 0 & 1 & 0%
\end{array}
\right) \varepsilon 
\]

\[
+\left( 
\begin{array}{llllll}
0 & 0 & 1 & 0 & 0 & 0 \\ 
0 & 1 & 0 & 1 & 0 & 0 \\ 
1 & 0 & 1 & 0 & 0 & 0 \\ 
0 & 1 & 0 & 0 & 0 & 0 \\ 
0 & 0 & 0 & 0 & 1 & 1 \\ 
0 & 0 & 0 & 0 & 1 & 0%
\end{array}
\right) \varepsilon ^{\prime }+\left( 
\begin{array}{llllll}
0 & 0 & 0 & 1 & 0 & 0 \\ 
0 & 0 & 1 & 0 & 0 & 0 \\ 
0 & 1 & 0 & 0 & 0 & 0 \\ 
1 & 0 & 0 & 0 & 0 & 0 \\ 
0 & 0 & 0 & 0 & 1 & 0 \\ 
0 & 0 & 0 & 0 & 0 & 1%
\end{array}
\right) \varepsilon ^{\prime \prime } 
\]

\begin{equation}
+\left( 
\begin{array}{llllll}
0 & 0 & 0 & 0 & 1 & 0 \\ 
0 & 0 & 0 & 0 & 1 & 1 \\ 
0 & 0 & 0 & 0 & 1 & 1 \\ 
0 & 0 & 0 & 0 & 1 & 0 \\ 
1 & 1 & 1 & 1 & 0 & 0 \\ 
0 & 1 & 1 & 0 & 0 & 0%
\end{array}
\right) \sigma +\left( 
\begin{array}{llllll}
0 & 0 & 0 & 0 & 0 & 1 \\ 
0 & 0 & 0 & 0 & 1 & 0 \\ 
0 & 0 & 0 & 0 & 1 & 0 \\ 
0 & 0 & 0 & 0 & 0 & 1 \\ 
0 & 1 & 1 & 0 & 0 & 0 \\ 
1 & 0 & 0 & 1 & 0 & 0%
\end{array}
\right) \sigma ^{\prime }  \label{/7}
\end{equation}

Using the equations (\ref{partion function1}) and (\ref{coefficients de
fusion}) we obtain the twelve following partition functions

\begin{eqnarray}
\mathcal{Z}_{I\mathcal{j}I} &=&\mathcal{Z}_{\varepsilon ^{\prime \prime }%
\mathcal{j}\varepsilon ^{\prime \prime }}=\chi _{1,1}(q),  \nonumber \\
\mathcal{Z}_{I\mathcal{j}\varepsilon } &=&\mathcal{Z}_{\varepsilon ^{\prime }%
\mathcal{j}\varepsilon ^{\prime \prime }}=\chi _{1,2}(q),  \nonumber \\
\mathcal{Z}_{I\mathcal{j}\varepsilon ^{\prime }} &=&\mathcal{Z}_{\varepsilon 
\mathcal{j}\varepsilon ^{\prime \prime }}=\chi _{1,3}(q),  \nonumber \\
\mathcal{Z}_{I\mathcal{j}\varepsilon ^{\prime \prime }} &=&\chi _{1,4}(q), 
\nonumber \\
\mathcal{Z}_{I\mathcal{j}\sigma } &=&\mathcal{Z}_{\varepsilon \mathcal{j}%
\sigma ^{\prime }}=\mathcal{Z}_{\varepsilon ^{\prime }\mathcal{j}\sigma
^{\prime }}=\mathcal{Z}_{\varepsilon ^{\prime \prime }\mathcal{j}\sigma
}=\chi _{2,2}(q),  \nonumber \\
\mathcal{Z}_{I\mathcal{j}\sigma ^{\prime }} &=&\mathcal{Z}_{\varepsilon
^{\prime \prime }\mathcal{j}\sigma ^{\prime }}=\chi _{2,4}(q),  \nonumber \\
\mathcal{Z}_{\varepsilon \mathcal{j}\varepsilon } &=&\mathcal{Z}%
_{\varepsilon ^{\prime }\mathcal{j}\varepsilon ^{\prime }}=\chi
_{1,1}(q)+\chi _{1,3}(q),  \nonumber \\
\mathcal{Z}_{\varepsilon \mathcal{j}\varepsilon ^{\prime }} &=&\chi
_{1,2}(q)+\chi _{1,4}(q),  \nonumber \\
\mathcal{Z}_{\varepsilon \mathcal{j}\sigma } &=&\mathcal{Z}_{\varepsilon
^{\prime }\mathcal{j}\sigma }=\chi _{2,2}(q)+\chi _{2,4}(q),  \nonumber \\
\mathcal{Z}_{\sigma \mathcal{j}\sigma } &=&\chi _{1,1}(q)+\chi
_{1,2}(q)+\chi _{1,3}(q)+\chi _{1,4}(q),  \nonumber \\
\mathcal{Z}_{\sigma \mathcal{j}\sigma ^{\prime }} &=&\chi _{1,2}(q)+\chi
_{1,3}(q),  \nonumber \\
\mathcal{Z}_{\sigma ^{\prime }\mathcal{j}\sigma ^{\prime }} &=&\chi
_{1,1}(q)+\chi _{1,4}(q)  \label{/8}
\end{eqnarray}
where we have taken into account the fact that 
\begin{equation}
\mathcal{Z}_{\alpha \mathcal{j}\beta }=\mathcal{Z}_{\beta \mathcal{j}\alpha }
\label{/9}
\end{equation}

\section{The boundary correlation functions}

\subsection{The coulomb gas formalism and the coherent boundary states}

In the coulomb gas-formalism one reproduces the primary fields $\phi _{r,s}$ 
$(0<r<p,0<s<p^{\prime })$ of the Kac table of the minimal model $%
(A_{p-1,}A_{p^{\prime }-1})$ by the vertex operators 
\begin{equation}
\mathcal{V}_{\alpha _{r,s}}(z)=:\exp (i\sqrt{2}\alpha _{r,s}\varphi (z):
\label{vertexoperator}
\end{equation}
with conformal dimensions given by 
\begin{equation}
h_{r,s}=\frac{1}{4}(r\alpha _{+}+s\alpha _{-})^{2}-\alpha _{0}^{2}
\end{equation}
where $\varphi (z)$ is the holomorphic part of a boson field $\Phi (z,%
\overline{z})$ and 
\begin{eqnarray}
\alpha _{r,s} &=&\frac{1}{2}(1-r)\alpha _{+}+\frac{1}{2}(1-s)\alpha
_{-}\qquad  \nonumber \\
\qquad \alpha _{+} &=&\sqrt{p/p^{\prime }}\qquad  \nonumber \\
\text{ \ }\alpha _{-} &=&-\sqrt{p^{\prime }/p}
\end{eqnarray}
The coefficient $\alpha _{0}$ is related to the central charge by 
\begin{equation}
c=1-24\alpha _{0}^{2}
\end{equation}
It is known (see for example \cite{Difrancesco}) that the correlation
function of $n$ vertex operators is equal to 
\begin{equation}
\left\langle \prod_{i=1}^{n}\mathcal{V}_{\alpha _{i}}(z_{i})\right\rangle
=\left\{ 
\begin{array}{c}
\prod_{i<j}^{n}(z_{i}-z_{j})^{2\alpha _{i}\alpha _{j}}\;\;\;\;\;\text{if \ \
\ \ \ \ \ \ }\sum_{i=1}^{n}\alpha _{i}=2\alpha _{0} \\ 
\\ 
0\;\ \ \ \ \ \ \ \ \ \ \ \ \ \ \ \ \ \ \ \ \ \ \ \text{otherwise}%
\end{array}
\right\}  \label{neutrality}
\end{equation}

The fock space $F_{\alpha ,\alpha _{0}}$ related to the vertex operator $%
\mathcal{V}_{\alpha }$ is built on the highest weight state 
\[
\left\vert \alpha ,\alpha _{0}\right\rangle =\exp (i\sqrt{2}\alpha \widehat{%
\varphi }_{0})\left\vert 0,\alpha _{0}\right\rangle 
\]%
by the action of the mode operators $\widehat{a}_{m}$ of the field $\varphi
(z)$, defined by 
\begin{equation}
\varphi (z)=\widehat{\varphi _{0}}-i\widehat{a}_{0}\ln (z)+i\sum_{n\neq 0}%
\frac{\widehat{a}_{n}}{n}z^{-n}
\end{equation}%
and satisfying the algebra 
\begin{eqnarray}
\left[ \widehat{a}_{m},\widehat{a}_{n}\right] &=&m\delta _{m+n} \\
\left[ \widehat{\varphi }_{0},\widehat{a}_{0}\right] &=&i  \nonumber
\end{eqnarray}%
The state $\left\vert 0,\alpha _{0}\right\rangle $ is the vacuum state and
we have 
\[
\widehat{a}_{0}\left\vert \alpha ,\alpha _{0}\right\rangle =\sqrt{2}\alpha
\left\vert \alpha ,\alpha _{0}\right\rangle 
\]%
\[
\left\langle \alpha ,\alpha _{0}|\beta ,\alpha _{0}\right\rangle =\kappa
\delta _{\alpha \beta } 
\]%
$\kappa $ being a normalization constant which can be chosen to be equal to
one.

The aim goal now is to construct conformally boundary states in terms of the
states $\left\vert \alpha ,\alpha _{0}\right\rangle $ of the different fock
spaces corresponding to the different values of $\alpha .$ Starting from the
ansatz 
\begin{equation}
\left\vert B_{\alpha ,\overline{\alpha },\alpha _{0}}\right\rangle
=\prod_{k>0}\exp (-\frac{1}{k}\widehat{a}_{-k}\widehat{\overline{a}}%
_{-k})\left\vert \alpha ,\overline{\alpha },\alpha _{0}\right\rangle
\end{equation}%
where the states $\left\vert \alpha ,\overline{\alpha },\alpha
_{0}\right\rangle $ are the direct product of the holomorphic and
antiholomorphic highest weight states 
\begin{equation}
\left\vert \alpha ,\overline{\alpha },\alpha _{0}\right\rangle =\left\vert
\alpha ,\alpha _{0}\right\rangle \otimes \left\vert \overline{\alpha }%
,\alpha _{0}\right\rangle
\end{equation}%
and by expressing the elements of the Virasoro algebra in terms of the
operators $\widehat{a}_{m}$, one can easily show that the states $\left\vert
B_{\alpha ,\overline{\alpha },\alpha _{0}}\right\rangle $ verify 
\begin{equation}
(L_{n}-\overline{L}_{-n})\left\vert B_{\alpha ,\overline{\alpha },\alpha
_{0}}\right\rangle =0  \label{Bstate}
\end{equation}%
only if 
\begin{equation}
\alpha +\overline{\alpha }-2\alpha _{0}=0
\end{equation}%
The boundary states satisfying these two conditions will be noted 
\begin{equation}
\left\vert B_{\alpha ,\overline{\alpha },\alpha _{0}}\right\rangle
=\left\vert B_{\alpha ,2\alpha _{0}-\alpha ,\alpha _{0}}\right\rangle
=\left\vert B(\alpha )\right\rangle
\end{equation}%
In addition to the necessary condition (\ref{Bstate}), the coherent states $%
\left\vert B(\alpha )\right\rangle $ have to satisfy the conditions obtained
from the duality of the partition function as already done in section 1.
There is in fact a combination of these states which satisfy the duality
condition and it is shown in \cite{shin1} that the Ishibashi states are
related to these states as 
\begin{equation}
|(r,s)>>=|a_{r,s}>+|a_{r,-s}>  \label{physcoherent}
\end{equation}%
with 
\begin{equation}
|a_{r,s}>=\sum_{k\in Z}|B(k\sqrt{pp^{\prime }}+\alpha _{r,s})>
\end{equation}

Using (\ref{physcoherent}) and (\ref{17}) one can write the physical (or
consistent) boundary states of any minimal model in terms of the coherent
ones. For the particular case of the minimal model $(A_{4},A_{3})$ we have
the following relations

\medskip

$\left| I\right\rangle =\frac{1}{\sqrt{2}}\frac{1}{\sqrt[4]{5}}\{\sqrt{s_{2}}%
(\left| a_{1,1}\right\rangle +\left| a_{1,-1}\right\rangle +\left|
a_{3,4}\right\rangle +\left| a_{3,-4}\right\rangle )+\sqrt{s_{1}}(\left|
a_{1,2}\right\rangle +$

$\qquad \left| a_{1,-2}\right\rangle +\left| a_{3,3}\right\rangle +\left|
a_{3,-3}\right\rangle )+\sqrt{s_{1}}(\left| a_{1,3}\right\rangle +\left|
a_{1,-3}\right\rangle +\left| a_{3,2}\right\rangle +\left|
a_{3,-2}\right\rangle )+$

$\qquad \sqrt{s_{2}}(\left| a_{1,4}\right\rangle +\left|
a_{1,-4}\right\rangle +\left| a_{3,1}\right\rangle +\left|
a_{3,-1}\right\rangle )+\sqrt{\sqrt{2}s_{1}}(\left| a_{2,2}\right\rangle
+\left| a_{2,-2}\right\rangle +$

$\qquad \left| a_{2,3}\right\rangle +\left| a_{2,-3}\right\rangle )+\sqrt{%
\sqrt{2}s_{2}}(\left| a_{2,4}\right\rangle +\left| a_{2,-4}\right\rangle
+\left| a_{2,1}\right\rangle +\left| a_{2,-1}\right\rangle )\}$

\medskip

$\left| \varepsilon \right\rangle =\frac{1}{\sqrt{2}}\frac{1}{\sqrt[4]{5}}\{%
\frac{s_{1}}{\sqrt{s_{2}}}(\left| a_{1,1}\right\rangle +\left|
a_{1,-1}\right\rangle +\left| a_{3,4}\right\rangle +\left|
a_{3,-4}\right\rangle )-\frac{s_{2}}{\sqrt{s_{1}}}(\left|
a_{1,2}\right\rangle +$

$\qquad \left| a_{1,-2}\right\rangle +\left| a_{3,3}\right\rangle +\left|
a_{3,-3}\right\rangle )-\frac{s_{2}}{\sqrt{s_{1}}}(\left|
a_{1,3}\right\rangle +\left| a_{1,-3}\right\rangle +\left|
a_{3,2}\right\rangle +\left| a_{3,-2}\right\rangle )+$

$\qquad \frac{s_{1}}{\sqrt{s_{2}}}(\left| a_{1,4}\right\rangle +\left|
a_{1,-4}\right\rangle +\left| a_{3,1}\right\rangle +\left|
a_{3,-1}\right\rangle )+\sqrt[4]{2}\frac{s_{2}}{\sqrt{s_{1}}}(\left|
a_{2,2}\right\rangle +\left| a_{2,-2}\right\rangle +$

$\qquad \left| a_{2,3}\right\rangle +\left| a_{2,-3}\right\rangle )-\sqrt[4]{%
2}\frac{s_{1}}{\sqrt{s_{2}}}(\left| a_{2,4}\right\rangle +\left|
a_{2,-4}\right\rangle +\left| a_{2,1}\right\rangle +\left|
a_{2,-1}\right\rangle )\}$

\medskip

$\left| \varepsilon ^{\prime }\right\rangle =\frac{1}{\sqrt{2}}\frac{1}{\sqrt%
[4]{5}}\{\frac{s_{1}}{\sqrt{s_{2}}}(\left| a_{1,1}\right\rangle +\left|
a_{1,-1}\right\rangle +\left| a_{3,4}\right\rangle +\left|
a_{3,-4}\right\rangle )-\frac{s_{2}}{\sqrt{s_{1}}}(\left|
a_{1,2}\right\rangle +$

$\qquad \left| a_{1,-2}\right\rangle +\left| a_{3,3}\right\rangle +\left|
a_{3,-3}\right\rangle )-\frac{s_{2}}{\sqrt{s_{1}}}(\left|
a_{1,3}\right\rangle +\left| a_{1,-3}\right\rangle +\left|
a_{3,2}\right\rangle +\left| a_{3,-2}\right\rangle )+$

$\qquad \frac{s_{1}}{\sqrt{s_{2}}}(\left| a_{1,4}\right\rangle +\left|
a_{1,-4}\right\rangle +\left| a_{3,1}\right\rangle +\left|
a_{3,-1}\right\rangle )-\sqrt[4]{2}\frac{s_{2}}{\sqrt{s_{1}}}(\left|
a_{2,2}\right\rangle +\left| a_{2,-2}\right\rangle +$

$\qquad \left| a_{2,3}\right\rangle +\left| a_{2,-3}\right\rangle )+\sqrt[4]{%
2}\frac{s_{1}}{\sqrt{s_{2}}}(\left| a_{2,4}\right\rangle +\left|
a_{2,-4}\right\rangle +\left| a_{2,1}\right\rangle +\left|
a_{2,-1}\right\rangle )\}$

\medskip

$\left| \varepsilon ^{\prime \prime }\right\rangle =\frac{1}{\sqrt{2}}\frac{1%
}{\sqrt[4]{5}}\{\sqrt{s_{2}}(\left| a_{1,1}\right\rangle +\left|
a_{1,-1}\right\rangle +\left| a_{3,4}\right\rangle +\left|
a_{3,-4}\right\rangle )+\sqrt{s_{1}}(\left| a_{1,2}\right\rangle +$

$\qquad \left| a_{1,-2}\right\rangle +\left| a_{3,3}\right\rangle +\left|
a_{3,-3}\right\rangle )+\sqrt{s_{1}}(\left| a_{1,3}\right\rangle +\left|
a_{1,-3}\right\rangle +\left| a_{3,2}\right\rangle +\left|
a_{3,-2}\right\rangle )+$

$\qquad \sqrt{s_{2}}(\left| a_{1,4}\right\rangle +\left|
a_{1,-4}\right\rangle +\left| a_{3,1}\right\rangle +\left|
a_{3,-1}\right\rangle )-\sqrt{\sqrt{2}s_{1}}(\left| a_{2,2}\right\rangle
+\left| a_{2,-2}\right\rangle +$

$\qquad \left| a_{2,3}\right\rangle +\left| a_{2,-3}\right\rangle )-\sqrt{%
\sqrt{2}s_{2}}(\left| a_{2,4}\right\rangle +\left| a_{2,-4}\right\rangle
+\left| a_{2,1}\right\rangle +\left| a_{2,-1}\right\rangle )\}$

\medskip

$\left| \sigma \right\rangle =\frac{1}{\sqrt[4]{5}}\{\frac{s_{1}}{\sqrt{s_{2}%
}}(\left| a_{1,1}\right\rangle +\left| a_{1,-1}\right\rangle +\left|
a_{3,4}\right\rangle +\left| a_{3,-4}\right\rangle )+\frac{s_{2}}{\sqrt{s_{1}%
}}(\left| a_{1,2}\right\rangle +$

$\qquad \left| a_{1,-2}\right\rangle +\left| a_{3,3}\right\rangle +\left|
a_{3,-3}\right\rangle )-\frac{s_{2}}{\sqrt{s_{1}}}(\left|
a_{1,3}\right\rangle +\left| a_{1,-3}\right\rangle +\left|
a_{3,2}\right\rangle +\left| a_{3,-2}\right\rangle )-$

$\qquad \frac{s_{1}}{\sqrt{s_{2}}}(\left| a_{1,4}\right\rangle +\left|
a_{1,-4}\right\rangle +\left| a_{3,1}\right\rangle +\left|
a_{3,-1}\right\rangle )$

\medskip

$\left| \sigma ^{\prime }\right\rangle =\frac{1}{\sqrt[4]{5}}\{\sqrt{s_{2}}%
(\left| a_{1,1}\right\rangle +\left| a_{1,-1}\right\rangle +\left|
a_{3,4}\right\rangle +\left| a_{3,-4}\right\rangle )-\sqrt{s_{1}}(\left|
a_{1,2}\right\rangle +$

$\qquad \left| a_{1,-2}\right\rangle +\left| a_{3,3}\right\rangle +\left|
a_{3,-3}\right\rangle )+\sqrt{s_{1}}(\left| a_{1,3}\right\rangle +\left|
a_{1,-3}\right\rangle +\left| a_{3,2}\right\rangle +\left|
a_{3,-2}\right\rangle )-$

$\qquad \sqrt{s_{2}}(\left| a_{1,4}\right\rangle +\left|
a_{1,-4}\right\rangle +\left| a_{3,1}\right\rangle +\left|
a_{3,-1}\right\rangle )$%
\begin{equation}  \label{/11}
\end{equation}

Note that before using (\ref{physcoherent}), we have replaced each state $%
|(r,s)\rangle \rangle $ by the symmetrised one 
\[
\frac{1}{\sqrt{2}}\left\{ |(r,s)\rangle \rangle +|(p^{\prime }-r,p-s)\rangle
\rangle \right\} 
\]

\subsection{The boundary 1-point correlation functions}

In the coulomb gas formalism a boundary p-points correlation function of the
form 
\[
\left\langle \alpha \right\vert \phi _{r_{1},s_{1}}(z_{1},\overline{z}%
_{1})\phi _{r_{2},s_{2}}(z_{2},\overline{z}_{2})......\phi
_{r_{p},s_{p}}(z_{p},\overline{z}_{p})\left\vert 0\right\rangle 
\]%
where $\left\vert \alpha \right\rangle $ is one of the allowed physical
boundary state for the model, can be written as a combination of correlators
having the general form 
\begin{equation}
\left\langle B(\alpha )\right\vert \prod_{i=1}^{p}\mathcal{V}%
_{(r_{i},s_{i}),(\overline{r}_{i},\overline{s}_{i})}^{(m_{i},n_{i}),(%
\overline{m}_{i},\overline{n}_{i})}(z_{i},\overline{z}_{i})\left\vert
0,0;\alpha _{0}\right\rangle  \label{boundarycorre}
\end{equation}%
Where we have noted 
\begin{equation}
\mathcal{V}_{(r_{i},s_{i}),(\overline{r}_{i},\overline{s}%
_{i})}^{(m_{i},n_{i}),(\overline{m}_{i},\overline{n}_{i})}(z_{i},\overline{z}%
_{i})=\mathcal{V}_{(r_{i},s_{i})}^{(m_{i},n_{i})}(z_{i})\overline{\mathcal{V}%
}_{(\overline{r}_{i},\overline{s}_{i})}^{(\overline{m}_{i},\overline{n}%
_{i})}(\overline{z}_{i})
\end{equation}%
with the screened vertex operators

\begin{eqnarray}
\mathcal{V}_{(r,s)}^{(m,n)}(z) &=&\oint
\prod_{i=1}^{m}du_{i}\prod_{j=1}^{n}dv_{j}\mathcal{V}_{r,s}(z)\mathcal{V}%
_{+}(u_{1})...\mathcal{V}_{+}(u_{m})\times \mathcal{V}_{-}(v_{1})...\mathcal{%
V}_{-}(v_{n}),  \nonumber \\
\overline{\mathcal{V}}_{(\overline{r},\overline{s})}^{(\overline{m},%
\overline{n})}(\overline{z}) &=&\oint \prod_{i=1}^{\overline{m}}d\overline{u}%
_{i}\prod_{j=1}^{\overline{n}}d\overline{v}_{j}\overline{\mathcal{V}}_{%
\overline{r},\overline{s}}(\overline{z})\mathcal{V}_{+}(\overline{u}_{1})...%
\mathcal{V}_{+}(\overline{u}_{\overline{m}})\times \mathcal{V}_{-}(\overline{%
v}_{1})...\mathcal{V}_{-}(\overline{v}_{\overline{n}})  \nonumber \\
&&  \label{88}
\end{eqnarray}
We recall that the numbers $m$ and $n$ of the screening operators 
\[
Q_{\pm }=\oint dz\mathcal{V}_{\alpha _{\pm }}(z)=\oint dz\mathcal{V}_{\pm
}(z) 
\]
(of conformal dimensions $1$ ) is so that the neutrality condition appearing
in (\ref{neutrality}) is fullfiled. For the correlator (\ref{boundarycorre})
the neutrality condition for the holomorphic part is 
\begin{equation}
-\alpha +\sum_{i}\alpha _{r_{i},s_{i}}+\sum_{i}m_{i}\alpha
_{+}\sum_{i}n_{i}\alpha _{-}=0  \label{neutre1}
\end{equation}
and for the antiholomortphic one, we have 
\begin{equation}
\alpha -2\alpha _{0}+\sum_{i}\alpha _{\overline{r}_{i},\overline{s}%
_{i}}+\sum_{i}\overline{m}_{i}\alpha _{+}\sum_{i}\overline{n}_{i}\alpha
_{-}=0  \label{neutre2}
\end{equation}

For example for the 1-point correlation function 
\begin{equation}
\left\langle B(\alpha )\right| \phi _{(r,s;\overline{r},\overline{s})}(z,%
\overline{z})\left| 0,0;\alpha _{0}\right\rangle =\left\langle B(\alpha
)\right| \mathcal{V}_{(r,s)}^{(m,n)}(z)\overline{\mathcal{V}}_{(r,s)}^{(%
\overline{m},\overline{n})}(\overline{z})\left| 0,0;\alpha _{0}\right\rangle
\end{equation}

with $(\overline{r},\overline{s})=(r,s)$. The combination of the equations (%
\ref{neutre1}) and (\ref{neutre2}) leads to

\begin{equation}
m=\overline{m}=n=\overline{n}=0  \label{100}
\end{equation}
So that in this case there is no screening operators and one find

\begin{equation}
\alpha =\alpha _{r,s}  \label{101}
\end{equation}

Then we have 
\begin{eqnarray}
\left\langle B(\alpha _{r,s})\right| \Phi _{(r,s;\overline{r},\overline{s}%
)}(z,\overline{z})\left| 0,0;\alpha _{0}\right\rangle &=&\left\langle
B(\alpha _{r,s})\right| \mathcal{V}_{(r,s)}(z)\overline{\mathcal{V}}_{(%
\overline{r},\overline{s})}(\overline{z})\left| 0,0;\alpha _{0}\right\rangle
\nonumber \\
&=&(1-z\overline{z})^{-2h_{r,s}}  \label{102}
\end{eqnarray}

\medskip Applying these algorithm for the Tricritical Ising model defined on
the unit disk (UD) (i.e in the $\zeta -plane)$ one obtains

\begin{eqnarray}
\left\langle I\left| I(\zeta ,\overline{\zeta })\right| 0\right\rangle
&=&\left\langle I\left| \mathcal{V}_{1,1}(\zeta )\overline{\mathcal{V}}%
_{3,4}(\overline{\zeta })\right| 0,0;\alpha _{0}\right\rangle  \nonumber \\
&=&\frac{1}{\sqrt[4]{5}}\sqrt{\frac{s_{2}}{2}}\left\langle B(\alpha
_{1,1})\right| \mathcal{V}_{1,1}(\zeta )\overline{\mathcal{V}}_{3,4}(%
\overline{\zeta })\left| 0,0;\alpha _{0}\right\rangle  \nonumber \\
&=&\frac{1}{\sqrt[4]{5}}\sqrt{\frac{s_{2}}{2}}(1-\zeta \overline{\zeta }%
)^{-2h_{1,1}}=\frac{1}{\sqrt[4]{5}}\sqrt{\frac{s_{2}}{2}},  \nonumber \\
\left\langle \varepsilon \left| I(\zeta ,\overline{\zeta })\right|
0\right\rangle &=&\left\langle \varepsilon \mathcal{j}0\right\rangle =\frac{1%
}{\sqrt[4]{5}}\frac{s_{1}}{\sqrt{2s_{2}}},  \nonumber \\
\left\langle \varepsilon ^{\prime }\left| I(\zeta ,\overline{\zeta })\right|
0\right\rangle &=&\left\langle \varepsilon ^{\prime }\mathcal{j}%
0\right\rangle =\frac{1}{\sqrt[4]{5}}\frac{s_{1}}{\sqrt{2s_{2}}},  \nonumber
\\
\left\langle \varepsilon ^{\prime \prime }\left| I(\zeta ,\overline{\zeta }%
)\right| 0\right\rangle &=&\left\langle \varepsilon ^{\prime \prime }%
\mathcal{j}0\right\rangle =\frac{1}{\sqrt[4]{5}}\sqrt{\frac{s_{2}}{2}}, 
\nonumber \\
\left\langle \sigma \left| I(\zeta ,\overline{\zeta })\right| 0\right\rangle
&=&\left\langle \sigma \mathcal{j}0\right\rangle =\frac{1}{\sqrt[4]{5}}\frac{%
s_{1}}{\sqrt{s_{2}}},  \nonumber \\
\left\langle \sigma ^{\prime }\left| I(\zeta ,\overline{\zeta })\right|
0\right\rangle &=&\left\langle \sigma ^{\prime }\mathcal{j}0\right\rangle =%
\frac{1}{\sqrt[4]{5}}\sqrt{s_{2}}  \label{/12}
\end{eqnarray}
For the operator $\varepsilon $%
\begin{eqnarray}
\left\langle I\left| \varepsilon (\zeta ,\overline{\zeta })\right|
0\right\rangle &=&\left\langle I\left| \mathcal{V}_{1,2}(\zeta )\overline{%
\mathcal{V}}_{3,3}(\overline{\zeta })\right| 0,0;\alpha _{0}\right\rangle 
\nonumber \\
&=&\frac{1}{\sqrt[4]{5}}\sqrt{\frac{s_{1}}{2}}\left\langle B(\alpha
_{1,2})\right| \mathcal{V}_{1,2}(\zeta )\overline{\mathcal{V}}_{3,3}(%
\overline{\zeta })\left| 0,0;\alpha _{0}\right\rangle  \nonumber \\
&=&\frac{1}{\sqrt[4]{5}}\sqrt{\frac{s_{1}}{2}}(1-\zeta \overline{\zeta })^{-%
\frac{1}{5}},  \nonumber \\
\left\langle \varepsilon \left| \varepsilon (\zeta ,\overline{\zeta }%
)\right| 0\right\rangle &=&-\frac{1}{\sqrt[4]{5}}\frac{s_{2}}{\sqrt{2s_{1}}}%
(1-\zeta \overline{\zeta })^{-\frac{1}{5}},  \nonumber \\
\left\langle \varepsilon ^{\prime }\left| \varepsilon (\zeta ,\overline{%
\zeta })\right| 0\right\rangle &=&-\frac{1}{\sqrt[4]{5}}\frac{s_{2}}{\sqrt{%
2s_{1}}}(1-\zeta \overline{\zeta })^{-\frac{1}{5}},  \nonumber \\
\left\langle \varepsilon ^{\prime \prime }\left| \varepsilon (\zeta ,%
\overline{\zeta })\right| 0\right\rangle &=&\frac{1}{\sqrt[4]{5}}\sqrt{\frac{%
s_{1}}{2}}(1-\zeta \overline{\zeta })^{-\frac{1}{5}},  \nonumber \\
\left\langle \sigma \left| \varepsilon (\zeta ,\overline{\zeta })\right|
0\right\rangle &=&\frac{1}{\sqrt[4]{5}}\frac{s_{2}}{\sqrt{s_{1}}}(1-\zeta 
\overline{\zeta })^{-\frac{1}{5}},  \nonumber \\
\left\langle \sigma ^{\prime }\left| \varepsilon (\zeta ,\overline{\zeta }%
)\right| 0\right\rangle &=&-\frac{1}{\sqrt[4]{5}}\sqrt{s_{1}}(1-\zeta 
\overline{\zeta })^{-\frac{1}{5}}  \label{/13}
\end{eqnarray}
and for the spin operator $\sigma $

\begin{eqnarray}
\left\langle I\left| \sigma (\zeta ,\overline{\zeta })\right| 0\right\rangle
&=&\frac{1}{\sqrt[4]{5}}s_{1}(1-\zeta \overline{\zeta })^{-\frac{3}{40}}, 
\nonumber \\
\left\langle \varepsilon \left| \sigma (\zeta ,\overline{\zeta })\right|
0\right\rangle &=&\frac{1}{\sqrt[4]{10}}\frac{s_{2}}{\sqrt{s_{1}}}(1-\zeta 
\overline{\zeta })^{-\frac{3}{40}},  \nonumber \\
\left\langle \varepsilon ^{\prime }\left| \sigma (\zeta ,\overline{\zeta }%
)\right| 0\right\rangle &=&-\frac{1}{\sqrt[4]{10}}\frac{s_{2}}{\sqrt{s_{1}}}%
(1-\zeta \overline{\zeta })^{-\frac{3}{40}},  \nonumber \\
\left\langle \varepsilon ^{\prime \prime }\left| \sigma (\zeta ,\overline{%
\zeta })\right| 0\right\rangle &=&-\frac{1}{\sqrt[4]{10}}\sqrt{s_{1}}%
(1-\zeta \overline{\zeta })^{-\frac{3}{40}},  \nonumber \\
\left\langle \sigma \left| \sigma (\zeta ,\overline{\zeta })\right|
0\right\rangle &=&0,  \nonumber \\
\left\langle \sigma ^{\prime }\left| \sigma (\zeta ,\overline{\zeta }%
)\right| 0\right\rangle &=&0  \label{/14}
\end{eqnarray}

\medskip

At this level one can come back to the UHP by the global conformal
transformation

\begin{equation}
z=-iy_{0}\frac{\zeta -1}{\zeta +1}\qquad ,\qquad \overline{z}=iy_{0}\frac{%
\overline{\zeta }-1}{\overline{\zeta }+1}  \label{Globaltransformation}
\end{equation}%
which map the origin $\zeta =0$ to the point $z=iy_{0}$. The effect of this
transformation on $n$-point correlation functions is given by 
\begin{eqnarray*}
&<&\phi _{1}(z_{1},\overline{z}_{1})\phi _{2}(z_{2},\overline{z}%
_{2})......\phi _{n}(z_{p},\overline{z}_{p})>_{UHP} \\
&=&\prod_{i=1}^{n}\left( \frac{dz_{i}}{d\zeta _{i}}\right) ^{-h_{i}}\left( 
\frac{d\overline{z}_{i}}{d\overline{\zeta }_{i}}\right) ^{-h_{i}}<\phi
_{1}(\zeta _{1},\overline{\zeta }_{1})\phi _{2}(\zeta _{2},\overline{\zeta }%
_{2})......\phi _{n}(\zeta _{p},\overline{\zeta }_{p})>_{Disk} \\
&=&\prod_{i=1}^{n}\left( \frac{2y_{0}}{(\zeta _{i}+1)(\overline{\zeta }%
_{i}+1)}\right) ^{-2h_{i}}<\phi _{1}(\zeta _{1},\overline{\zeta }_{1})\phi
_{2}(\zeta _{2},\overline{\zeta }_{2})......\phi _{n}(\zeta _{p},\overline{%
\zeta }_{p})>_{Disk}
\end{eqnarray*}

\medskip

For example the one point correlators given in equations (\ref{/13}) and (%
\ref{/14})are transformed by the transformation (\ref{Globaltransformation})
into

\begin{eqnarray}
\left\langle \varepsilon (z,\overline{z})\right\rangle _{I} &=&\frac{%
\left\langle I\left| \varepsilon (z,\overline{z})\right| 0,0;\alpha
_{0}\right\rangle }{\left\langle I\mathcal{j}0,0;\alpha _{0}\right\rangle }%
=\left( \frac{s_{1}}{s_{2}}\right) ^{\frac{1}{2}}(2y)^{-\frac{1}{5}}, 
\nonumber \\
\left\langle \varepsilon (z,\overline{z})\right\rangle _{\varepsilon }
&=&-\left( \frac{s_{2}}{s_{1}}\right) ^{\frac{3}{2}}(2y)^{-\frac{1}{5}}, 
\nonumber \\
\left\langle \varepsilon (z,\overline{z})\right\rangle _{\varepsilon
^{\prime }} &=&-\left( \frac{s_{2}}{s_{1}}\right) ^{\frac{3}{2}}(2y)^{-\frac{%
1}{5}},  \nonumber \\
\left\langle \varepsilon (z,\overline{z})\right\rangle _{\varepsilon
^{\prime \prime }} &=&\left( \frac{s_{1}}{s_{2}}\right) ^{\frac{1}{2}}(2y)^{-%
\frac{1}{5}},  \nonumber \\
\left\langle \varepsilon (z,\overline{z})\right\rangle _{\sigma } &=&\left( 
\frac{s_{2}}{s_{1}}\right) ^{\frac{3}{2}}(2y)^{-\frac{1}{5}},  \nonumber \\
\left\langle \varepsilon (z,\overline{z})\right\rangle _{\sigma ^{\prime }}
&=&-\left( \frac{s_{1}}{s_{2}}\right) ^{\frac{1}{2}}(2y)^{-\frac{1}{5}}, 
\nonumber \\
\left\langle \sigma (z,\overline{z})\right\rangle _{I} &=&\sqrt{2}\frac{s_{1}%
}{\sqrt{s_{2}}}(2y)^{-\frac{3}{40}},  \nonumber \\
\left\langle \sigma (z,\overline{z})\right\rangle _{\varepsilon } &=&\sqrt[4]%
{2}\left( \frac{s_{2}}{s_{1}}\right) ^{\frac{3}{2}}(2y)^{-\frac{3}{40}}, 
\nonumber \\
\left\langle \sigma (z,\overline{z})\right\rangle _{\varepsilon ^{\prime }}
&=&-\sqrt[4]{2}\left( \frac{s_{2}}{s_{1}}\right) ^{\frac{3}{2}}(2y)^{-\frac{3%
}{40}},  \nonumber \\
\left\langle \sigma (z,\overline{z})\right\rangle _{\varepsilon ^{\prime
\prime }} &=&-\sqrt[4]{2}\left( \frac{s_{1}}{s_{2}}\right) ^{\frac{1}{2}%
}(2y)^{-\frac{3}{40}},  \nonumber \\
\left\langle \sigma (z,\overline{z})\right\rangle _{\sigma } &=&0,  \nonumber
\\
\left\langle \sigma (z,\overline{z})\right\rangle _{\sigma ^{\prime }} &=&0
\label{/16}
\end{eqnarray}

\subsection{The boundary 2-point boundary correlation functions:}

We consider the two point function of the field $\Phi _{1,2}\equiv
\varepsilon $. They are of the form 
\begin{eqnarray}
&&\left\langle B(\alpha )\right\vert \Phi _{1,2}(\zeta _{1},\overline{\zeta }%
_{1})\Phi _{1,2}(\zeta _{2},\overline{\zeta }_{2})\left\vert 0,0;\alpha
_{0}\right\rangle  \nonumber \\
&=&\left\langle B(\alpha )\right\vert \mathcal{V}_{1,2}^{m_{1},n_{1}}(\zeta
_{1})\overline{\mathcal{V}}_{(p^{\prime }-1,p-2)}^{\overline{m}_{1},%
\overline{n}_{1}}(\overline{\zeta }_{1})\mathcal{V}_{1,2}^{m_{2},n_{2}}(%
\zeta _{2})\overline{\mathcal{V}}_{1,2}^{\overline{m}_{2},\overline{n}_{2}}(%
\overline{\zeta }_{2})\left\vert 0,0;\alpha _{0}\right\rangle  \label{ff104}
\end{eqnarray}
where we taked first $\Phi _{1,2}(\zeta _{1},\overline{\zeta }_{1})\sim 
\mathcal{V}_{1,2}(\zeta _{1})\overline{\mathcal{V}}_{(p^{\prime }-1,p-2)}(%
\overline{\zeta }_{1})$ and then $\Phi _{1,2}(\zeta _{1},\overline{\zeta }%
_{1})\sim \mathcal{V}_{1,2}(\zeta _{2})\overline{\mathcal{V}}_{1,2}(%
\overline{\zeta }_{2}).$

From the neutrality condition, we obtain the following constraints

\begin{eqnarray}
\alpha &=&(m_{1}+m_{2})\alpha _{+}+(n_{1}+n_{2}-1)\alpha _{-},  \nonumber \\
2\alpha _{0}-\alpha &=&(\overline{m}_{1}+\overline{m}_{2}+1)\alpha _{+}+(%
\overline{n}_{1}+\overline{n}_{2}+1)\alpha _{-}  \label{105}
\end{eqnarray}
which can be solved as

\begin{eqnarray}
m &=&\overline{m}=\overline{n}=0,n=1,  \label{106a} \\
m &=&\overline{m}=n=0,\overline{n}=1  \label{106b}
\end{eqnarray}
with $m=m_{1}+m_{2},n=n_{1}+n_{2},\overline{m}=\overline{m}_{1}+\overline{m}%
_{2},\overline{n}=\overline{n}_{1}+\overline{n}_{2}.$

The first solution (\ref{106a}) corresponds to $\alpha =\alpha _{1,1}=0$,
and to the conformal block 
\begin{eqnarray}
I_{1} &=&N_{1}(1-\zeta _{1}\overline{\zeta }_{1})^{a}(1-\zeta _{2}\overline{%
\zeta }_{2})^{a}\left[ \eta (\eta -1)\right] ^{a}  \label{107} \\
&&\times \frac{\Gamma (1-\alpha _{-}^{2})^{2}}{\Gamma (2-2\alpha _{-}^{2})}%
F(2a,1-\alpha _{-}^{2},2-2\alpha _{-}^{2};\eta )
\end{eqnarray}
with $a=2\alpha _{1,2}(2\alpha _{0}-\alpha _{1,2})$, $F$ being the
hypergeometric function and $\eta $ is the parameter defined by 
\[
\eta =\frac{(\zeta _{1}-\zeta _{2})(\overline{\zeta }_{1}-\overline{\zeta }%
_{2})}{(1-\zeta _{1}\overline{\zeta }_{1})(1-\zeta _{2}\overline{\zeta }_{2})%
} 
\]

The second solution (\ref{106b}) gives $\alpha =\alpha _{1,3}=-\alpha _{-}$,
and corresponds to the conformal block

\begin{eqnarray}
I_{2} &=&N_{2}(1-\zeta _{1}\overline{\zeta }_{1})^{a}(1-\zeta _{2}\overline{%
\zeta }_{2})^{a}\left[ \eta (\eta -1)\right] ^{a}(-\eta )^{b-a}  \nonumber \\
&&\times \frac{\Gamma (1-\alpha _{-}^{2})\Gamma (3\alpha _{-}^{2}-1)}{\Gamma
(2\alpha _{-}^{2})}F(\alpha _{-}^{2},1-\alpha _{-}^{2},2\alpha _{-}^{2};\eta
)  \label{108}
\end{eqnarray}
with $b=2\alpha _{1,2}^{2}.$

Using (\ref{Globaltransformation}) one can write these equations on the UHP
as

\begin{eqnarray}
I_{1} &=&N_{1}\left\{ \frac{(z_{1}-\overline{z}_{1})(\overline{z}_{2}-z_{2})%
}{(z_{1}-z_{2})(\overline{z}_{1}-\overline{z}_{2})(z_{1}-\overline{z}_{2})(%
\overline{z}_{1}-z_{2})}\right\} ^{2h_{1,2}}  \nonumber \\
&&\times \frac{\Gamma (1-\alpha _{-}^{2})^{2}}{\Gamma (2-2\alpha _{-}^{2})}%
F(-4h_{1,2},1-\alpha _{-}^{2},2-2\alpha _{-}^{2};\eta )  \label{109}
\end{eqnarray}

and 
\begin{eqnarray}
I_{2} &=&N_{2}\left\{ \frac{(z_{1}-\overline{z}_{1})(\overline{z}_{2}-z_{2})%
}{(z_{1}-z_{2})(\overline{z}_{1}-\overline{z}_{2})(z_{1}-\overline{z}_{2})(%
\overline{z}_{1}-z_{2})}\right\} ^{2h_{1,2}}  \nonumber \\
&&\times \frac{\Gamma (1-\alpha _{-}^{2})\Gamma (3\alpha _{-}^{2}-1)}{\Gamma
(2\alpha _{-}^{2})}F(\alpha _{-}^{2},1-\alpha _{-}^{2},2\alpha _{-}^{2};\eta
)  \label{110}
\end{eqnarray}
with 
\[
\eta =\frac{(\zeta _{1}-\zeta _{2})(\overline{\zeta }_{1}-\overline{\zeta }%
_{2})}{(1-\zeta _{1}\overline{\zeta }_{1})(1-\zeta _{2}\overline{\zeta }_{2})%
}\rightarrow \frac{(z_{1}-z_{2})(\overline{z}_{1}-\overline{z}_{2})}{(z_{1}-%
\overline{z}_{1})(z_{2}-\overline{z}_{2})} 
\]

Combining the two blocks with the coefficients appearing in the development
of the boundary states we obtain for the correlators in the UHP 
\begin{eqnarray*}
\left\langle \varepsilon (z_{1},\overline{z}_{1})\varepsilon (z_{2},%
\overline{z}_{2})\right\rangle _{\widetilde{I}} &=&\left\{ \frac{(z_{1}-%
\overline{z}_{1})(\overline{z}_{2}-z_{2})}{(z_{1}-z_{2})(\overline{z}_{1}-%
\overline{z}_{2})(z_{1}-\overline{z}_{2})(\overline{z}_{1}-z_{2})}\right\} ^{%
\frac{1}{5}}\times \\
&&\left\{ N_{1}\frac{\Gamma (1/5)^{2}}{\Gamma (2/5)}F_{1}+\sqrt{\frac{s_{1}}{%
s_{2}}}N_{2}\frac{\Gamma (\frac{1}{5})\Gamma (\frac{7}{5})}{\Gamma (\frac{8}{%
5})}\times (-\eta )^{9/5}F_{2}\right\} ,
\end{eqnarray*}

\begin{eqnarray*}
\left\langle \varepsilon (z_{1},\overline{z}_{1})\varepsilon (z_{2},%
\overline{z}_{2})\right\rangle _{\widetilde{\varepsilon }} &=&\left\{ \frac{%
(z_{1}-\overline{z}_{1})(\overline{z}_{2}-z_{2})}{(z_{1}-z_{2})(\overline{z}%
_{1}-\overline{z}_{2})(z_{1}-\overline{z}_{2})(\overline{z}_{1}-z_{2})}%
\right\} ^{\frac{1}{5}}\times \\
&&\left\{ N_{1}\frac{\Gamma (1/5)^{2}}{\Gamma (2/5)}F_{1}-\left( \frac{s_{2}%
}{s_{1}}\right) ^{3/2}N_{2}\frac{\Gamma (\frac{1}{5})\Gamma (\frac{7}{5})}{%
\Gamma (\frac{8}{5})}\times (-\eta )^{9/5}F_{2}\right\} ,
\end{eqnarray*}

\begin{eqnarray*}
\left\langle \varepsilon (z_{1},\overline{z}_{1})\varepsilon (z_{2},%
\overline{z}_{2})\right\rangle _{\widetilde{\varepsilon }^{\prime }}
&=&\left\{ \frac{(z_{1}-\overline{z}_{1})(\overline{z}_{2}-z_{2})}{%
(z_{1}-z_{2})(\overline{z}_{1}-\overline{z}_{2})(z_{1}-\overline{z}_{2})(%
\overline{z}_{1}-z_{2})}\right\} ^{\frac{1}{5}}\times \\
&&\left\{ N_{1}\frac{\Gamma (1/5)^{2}}{\Gamma (2/5)}F_{1}-\left( \frac{s_{2}%
}{s_{1}}\right) ^{3/2}N_{2}\frac{\Gamma (\frac{1}{5})\Gamma (\frac{7}{5})}{%
\Gamma (\frac{8}{5})}\times (-\eta )^{9/5}F_{2}\right\} ,
\end{eqnarray*}

\begin{eqnarray*}
\left\langle \varepsilon (z_{1},\overline{z}_{1})\varepsilon (z_{2},%
\overline{z}_{2})\right\rangle _{\widetilde{\varepsilon }^{\prime \prime }}
&=&\left\{ \frac{(z_{1}-\overline{z}_{1})(\overline{z}_{2}-z_{2})}{%
(z_{1}-z_{2})(\overline{z}_{1}-\overline{z}_{2})(z_{1}-\overline{z}_{2})(%
\overline{z}_{1}-z_{2})}\right\} ^{\frac{1}{5}}\times \\
&&\left\{ N_{1}\frac{\Gamma (1/5)^{2}}{\Gamma (2/5)}F_{1}+\left( \frac{s_{1}%
}{s_{2}}\right) ^{1/2}N_{2}\frac{\Gamma (\frac{1}{5})\Gamma (\frac{7}{5})}{%
\Gamma (\frac{8}{5})}\times (-\eta )^{9/5}F_{2}\right\} ,
\end{eqnarray*}

\begin{eqnarray*}
\left\langle \varepsilon (z_{1},\overline{z}_{1})\varepsilon (z_{2},%
\overline{z}_{2})\right\rangle _{\widetilde{\sigma }} &=&\left\{ \frac{%
(z_{1}-\overline{z}_{1})(\overline{z}_{2}-z_{2})}{(z_{1}-z_{2})(\overline{z}%
_{1}-\overline{z}_{2})(z_{1}-\overline{z}_{2})(\overline{z}_{1}-z_{2})}%
\right\} ^{\frac{1}{5}}\times \\
&&\left\{ N_{1}\frac{\Gamma (1/5)^{2}}{\Gamma (2/5)}F_{1}-\left( \frac{s_{2}%
}{s_{1}}\right) ^{3/2}N_{2}\frac{\Gamma (\frac{1}{5})\Gamma (\frac{7}{5})}{%
\Gamma (\frac{8}{5})}\times (-\eta )^{9/5}F_{2}\right\} ,
\end{eqnarray*}

\begin{eqnarray*}
\left\langle \varepsilon (z_{1},\overline{z}_{1})\varepsilon (z_{2},%
\overline{z}_{2})\right\rangle _{\widetilde{\sigma }^{\prime }} &=&\left\{ 
\frac{(z_{1}-\overline{z}_{1})(\overline{z}_{2}-z_{2})}{(z_{1}-z_{2})(%
\overline{z}_{1}-\overline{z}_{2})(z_{1}-\overline{z}_{2})(\overline{z}%
_{1}-z_{2})}\right\} ^{\frac{1}{5}}\times \\
&&\left\{ N_{1}\frac{\Gamma (1/5)^{2}}{\Gamma (2/5)}F_{1}+\left( \frac{s_{1}%
}{s_{2}}\right) ^{1/2}N_{2}\frac{\Gamma (\frac{1}{5})\Gamma (\frac{7}{5})}{%
\Gamma (\frac{8}{5})}\times (-\eta )^{9/5}F_{2}\right\}
\end{eqnarray*}

\begin{equation}  \label{/17}
\end{equation}%
with $F_{1}=F(-\frac{4}{10},\frac{1}{5},\frac{2}{5};\eta ),$ et $F_{2}=F(%
\frac{4}{5},\frac{1}{5},\frac{8}{5};\eta )$.

\section{Conclusion}

We have obtained the\ boundary consistent states and the exact correlation
functions of the tricritical Ising model, using the Coulomb-gas formalism.
The results obtained for 1-point correlators are in concordance with that
already known (see for example chapter 15 of \cite{Henkel} ).

We hope also to generalize the use of this method to the case of the
nondiagonal minimal models, such as the one representing the tricritical
3-states Potts model.

\medskip

\textbf{Acknowledgments\newline
}

S.B. would like to thank T. Wydro and J.F. McCabe for introducing him to the
computational methods used in conformal field theories and statistical
physics, and T. Wydro for the warm hospitality during the stay at the
university Paul Verlaine of Metz.

\end{document}